\documentstyle[sprocl,epsfig]{article}

\begin{document}

\begin{center}
{\tiny Conference on Perspectives in Nuclear Physics at
Intermediate Energies, Trieste, 8-12 maggio 1995}
\end{center}
\vskip 1cm

\title{ PARITY-VIOLATING LONGITUDINAL RESPONSE
 }
\author{ A. DE PACE
 }
\address{ Istituto Nazionale di Fisica Nucleare, Sezione di Torino,
 via Giuria 1, \\ I-10125 Torino, Italy
 }

\maketitle\abstracts{
The parity-violating quasielastic electron scattering response is explored
within the context of a model that builds antisymmetrized random phase
approximation and Hartree-Fock correlations on a relativistic Fermi gas basis.
Particular emphasis is put on the weak-neutral longitudinal response function,
since this observable displays a strong sensitivity to isospin correlations:
specifically, it is shown how, through a diagrammatic cancellation/filtration
mechanism, this response acts as a magnifier of pionic correlations in the
nuclear medium. The parity-violating longitudinal response function also
displays appreciable sensitivity to the electric strangeness content of the
nucleon, thus making quasielastic electron scattering a possible candidate to
measure the nucleon electric strange form factor at relatively high momentum
transfers. Finally, we discuss how observables, related to the asymmetry, can
be
constructed to disentangle the nuclear and nucleonic effects.
 }

\section{ Introduction
 }
\label{sec:intro}

Parity-violating (pv) electron scattering is a promising tool to study the
structure of the nucleon, a notable issue being, for instance, the strangeness
content (for a general review of pv studies see Ref.~\cite{Mus94}).
The proton alone, however, is not enough to extract sufficiently constrained
information, because of the large number of poorly known form factors entering
the game. One is thus led to consider the neutron as well, which in turn
implies, generally, the use of nuclei as target. A natural first choice is then
elastic scattering off very light nuclei, in order to minimize the nuclear
structure uncertainties and to have a nuclear form factor that falls off not
too
rapidly with the momentum transfer.

At intermediate energies also quasielastic scattering shows promising features,
such as large cross sections (being proportional to the number of nucleons) and
nuclear form factors slowly decreasing with the momentum transfer.\cite{Don92}
The issue here is, of course, whether one is able to control the nuclear
dynamics sufficiently well to extract accurate information on the single
nucleon
form factors. This issue has been discussed at length in Ref.~\cite{Bar94},
where it has been shown that observables can be constructed, which are
selectively sensitive to the nuclear or nucleonic physics content (more on this
in the last Section).

In the following we focus on the pv (or weak-neutral) longitudinal response,
which had been shown to be extraordinarily sensitive to nuclear isospin
correlations~\cite{Alb93}; this response function also displays an appreciable
sensitivity to the nucleon electric strange form factor and, as mentioned
above,
in the last Section we discuss a possible way to disentangle the two
dependences.

Let us start by introducing the asymmetry ${\cal A}$, defined in terms of the
double-differential cross sections, $d^2\sigma^{\pm}/d\Omega d\epsilon'$, for
longitudinally polarized electrons with helicity $\pm1$ as~\cite{Don92}:
\begin{equation}
\begin{array}{rcl}
  {\cal A} &=&
  \frac{\strut d^2\sigma^+ - d^2\sigma^-}{\strut d^2\sigma^+ + d^2 \sigma^-} \\
           &=& {\cal A}_0
    \frac{\strut v_L R^L_{AV} (q,\omega) + v_T R^T_{AV} (q,\omega) +
     v_{T'} R^{T'}_{VA} (q,\omega)}{\strut v_L R^L(q,\omega) +
       v_T R^T(q,\omega) }\ ,
\end{array}
\label{eq:asymmetry}
\end{equation}
where $v_L$, $v_T$ and $v_T'$ are kinematical factors and
\begin{equation}
  {\cal{A}}_0 = {G\left| Q^2\right|\over 2\pi\alpha\sqrt{2}}\simeq
    3.1\times 10^{-4}\tau
\label{eq:A0}
\end{equation}
sets the scale of the asymmetry; here $\tau=|Q^2|/(4m_N^2)$
--- $Q^2=\omega^2-q^2$ being the four-momentum transfer, ---
$\alpha$ is the electromagnetic and $G$ the Fermi coupling constant.

In the denominator of Eq.~(\ref{eq:asymmetry}) the standard electromagnetic
(em)
response functions appear, while the numerator contains the new weak-neutral
longitudinal, transverse and axial responses, respectively.

\section{ Weak-neutral longitudinal response
 }
\label{sec:w-nlr}

\begin{figure}[p]
\begin{center}
\mbox{\epsfig{file=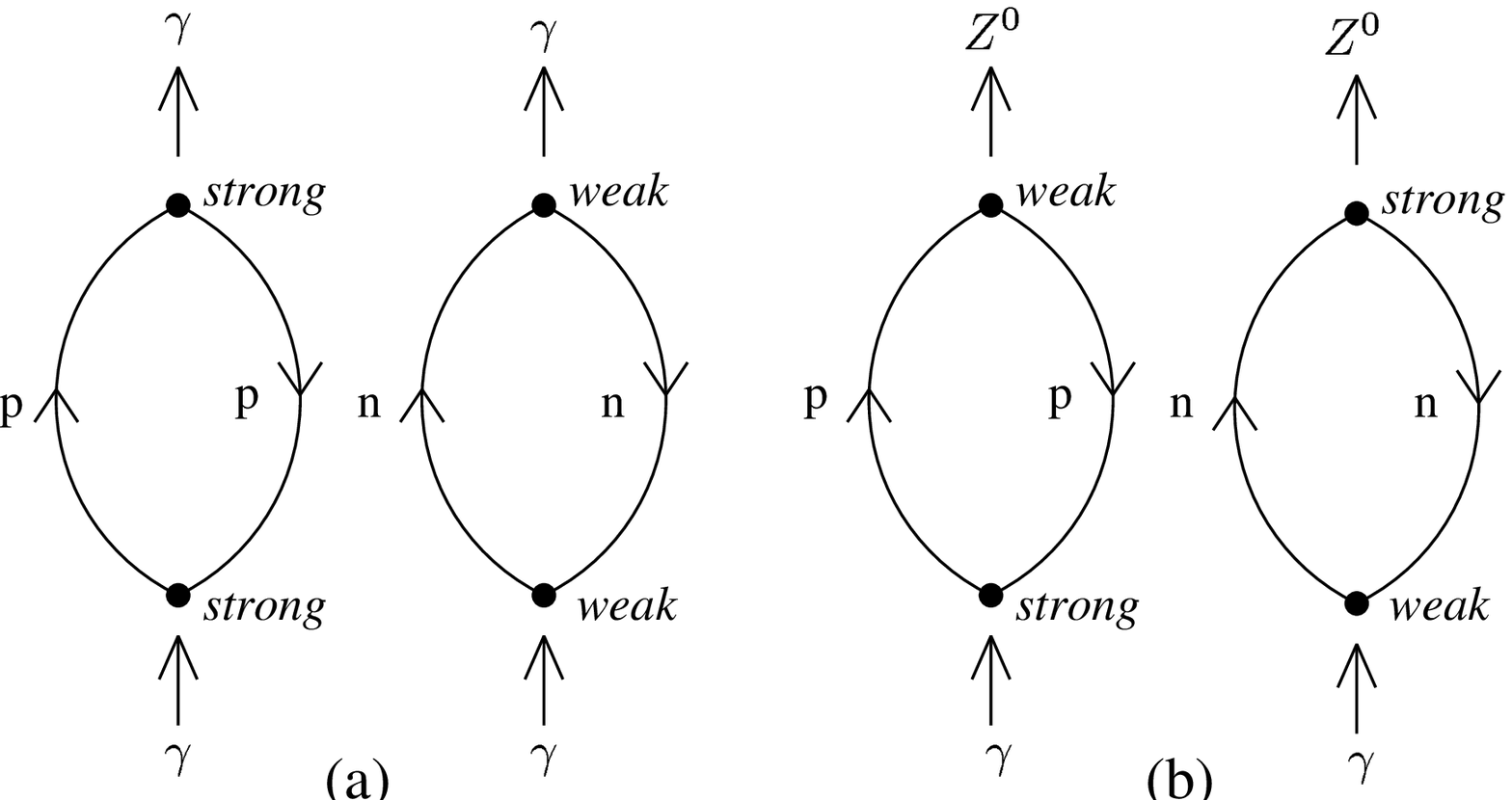,width=.75\textwidth,height=3.9cm}}
\vskip -3mm
\end{center}
\caption{ Feynman diagrams representing the free particle-hole polarization
propagator for the em (a) and pv (b) longitudinal response.
The excitation of proton (p) and neutron (n) particle-hole pairs is shown
separately. The labels {\em strong} and {\em weak} refer to the strength of
the nucleon coupling to the photon $\gamma$ or to the vector boson $Z^0$.
  }
\label{fig:diagrams-free}
\vfill
\vskip 4mm
\begin{center}
\mbox{\epsfig{file=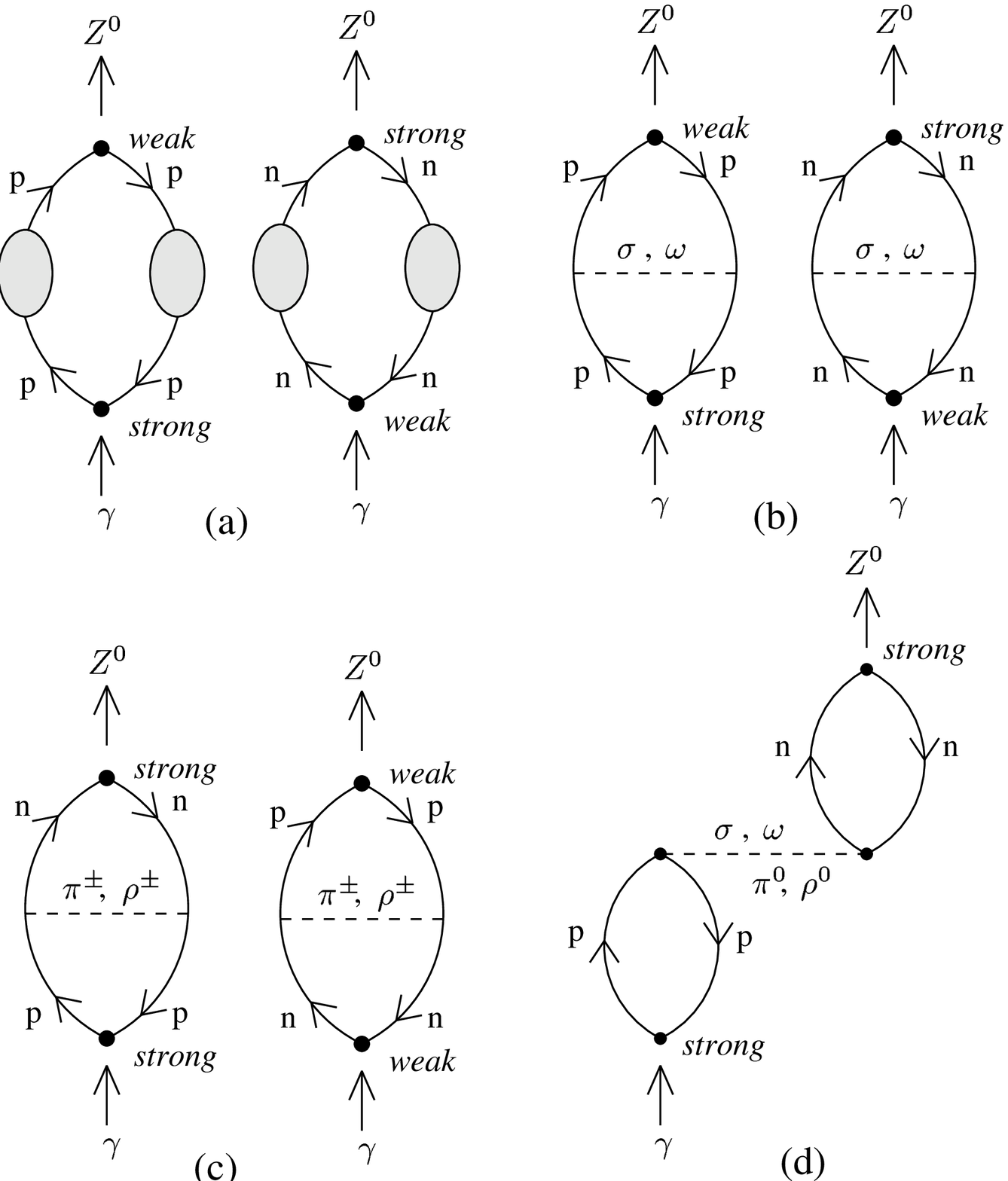,width=.75\textwidth}}
\vskip -3mm
\end{center}
\caption{ Feynman diagrams corresponding to the correlated pv longitudinal
response.
The meaning of the labels is as in Fig.~\protect{\ref{fig:diagrams-free}}.
In (a) one has the HF case, where the bubbles represent the nucleon
self-energy; in (b) and (c) the first-order exchange diagrams induced by the
exchange of isoscalar and isovector mesons, respectively;
in (d) a first-order ring diagram involving the exchange of neutral mesons.
  }
\label{fig:diagrams-corr}
\end{figure}

The peculiar role played by $R^L_{AV}$ can be understood by decomposing the
response functions into their isospin components.
The em charge response then reads
\begin{equation}
  R^L(q,\omega) = R^L(\tau=0) + R^L(\tau=1),
\label{eq:RL}
\end{equation}
with the isoscalar and isovector pieces entering with the same sign and the
same
norm in an independent-particle model (apart from effects due to the nucleon
form factors). The weak-neutral longitudinal response, on the other hand, is
given by
\begin{equation}
  R^L_{AV}(q,\omega) = - \left[\beta_V^{(0)} R^L(\tau=0) + \beta_V^{(1)}
    R^L(\tau=1)\right],
\label{eq:RLAV}
\end{equation}
where
\begin{equation}
\begin{array}{rcl}
  \beta_V^{(0)} &=& 1 - 2 \sin^2 \theta_W \\
  \beta_V^{(1)} &=& - 2 \sin^2 \theta_W .
\end{array}
\end{equation}
Since $\sin^2 \theta_W$ = 0.227 ($\theta_W$ being the Weinberg angle), one has
$\beta_V^{(0)}\approx -\beta_V^{(1)}$ and one gets a combination of the isospin
components that is nearly {\em orthogonal} to the one entering
Eq.~(\ref{eq:RL});
in particular, in an independent-particle model $R^L_{AV}$ nearly vanishes.
It is however clear that any nuclear correlations altering the
isoscalar-isovector balance will markedly affect $R^L_{AV}$.

A more direct picture of why $R^L_{AV}$ is so small in the independent-particle
model and why isospin correlations have dramatic effects (as we shall see
quantitatively later on) can be gained looking at the representation of the
response function in terms of many-body Feynman diagrams. Indeed, by inspecting
Fig.~\ref{fig:diagrams-free} one can easily understand why the uncorrelated em
response is expected to be larger than its pv counterpart, since the
corresponding polarization propagator can involve two strong $\gamma p$
vertices
(here {\em strong} and {\em weak} refer to the relative strength of the
couplings), while in the other case there must be a weak $\gamma n$ or
$Z^0 p$ vertex (note that the $Z^0$ couples strongly to the neutron and weakly
to the proton).

Introducing correlations, one sees (Fig.~\ref{fig:diagrams-corr}) that there
are
classes of diagrams that cannot change this picture, such as Hartree-Fock (HF)
insertions (Fig.~\ref{fig:diagrams-corr}a) or exchange contributions involving
neutral mesons (Fig.~\ref{fig:diagrams-corr}b). On the other hand, exchange
diagrams involving {\em charged} mesons (Fig.~\ref{fig:diagrams-corr}c) or ring
diagrams with neutral mesons (unless forbidden by spin selection rules, as in
the case of $\pi^0$ and $\rho^0$) (Fig.~\ref{fig:diagrams-corr}d) can turn a
proton into a neutron: because of the two strong vertices, these contributions
will now dominate over the free ones.

\subsection{ Nuclear correlations
}
\label{subsec:nuclcorr}

\begin{figure}[p]
\begin{center}
\mbox{\epsfig{file=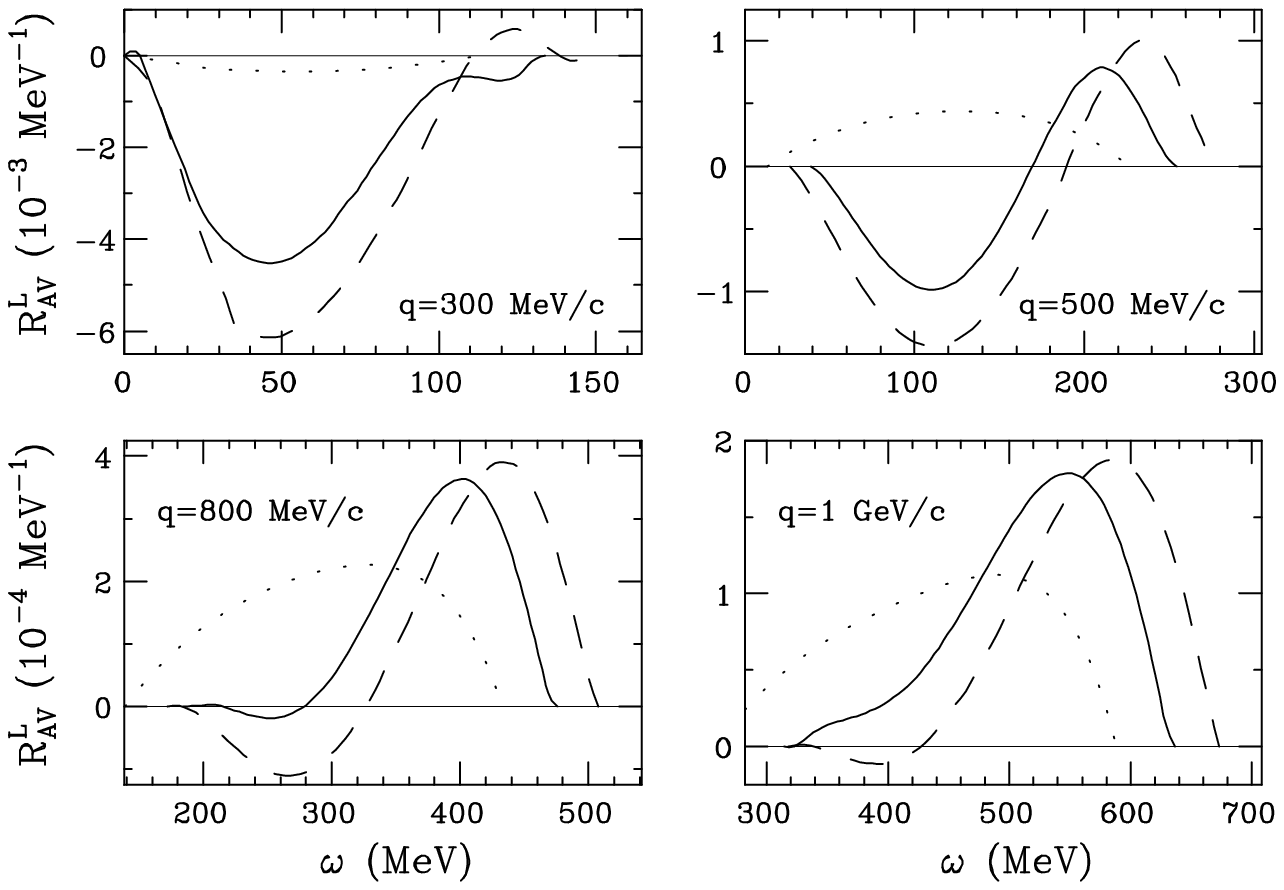,height=68mm}}
\vskip -3mm
\end{center}
\caption{
The pv longitudinal response $R^L_{AV}$ is shown as a function of
$\omega$ at $q=$ 300, 500, 800 and 1000 MeV/c. The dotted curves correspond
to the free RFG calculation with $k_F=$ 225 MeV/c, the dashed curves to the
HF-RPA calculation with $k_F=$ 225 MeV/c, the solid curves to HF-RPA with
$k_F=$ 200 MeV/c.
  }
\label{fig:RLAV-HF-RPA}
\vfill
\vskip 3mm
\begin{center}
\mbox{\epsfig{file=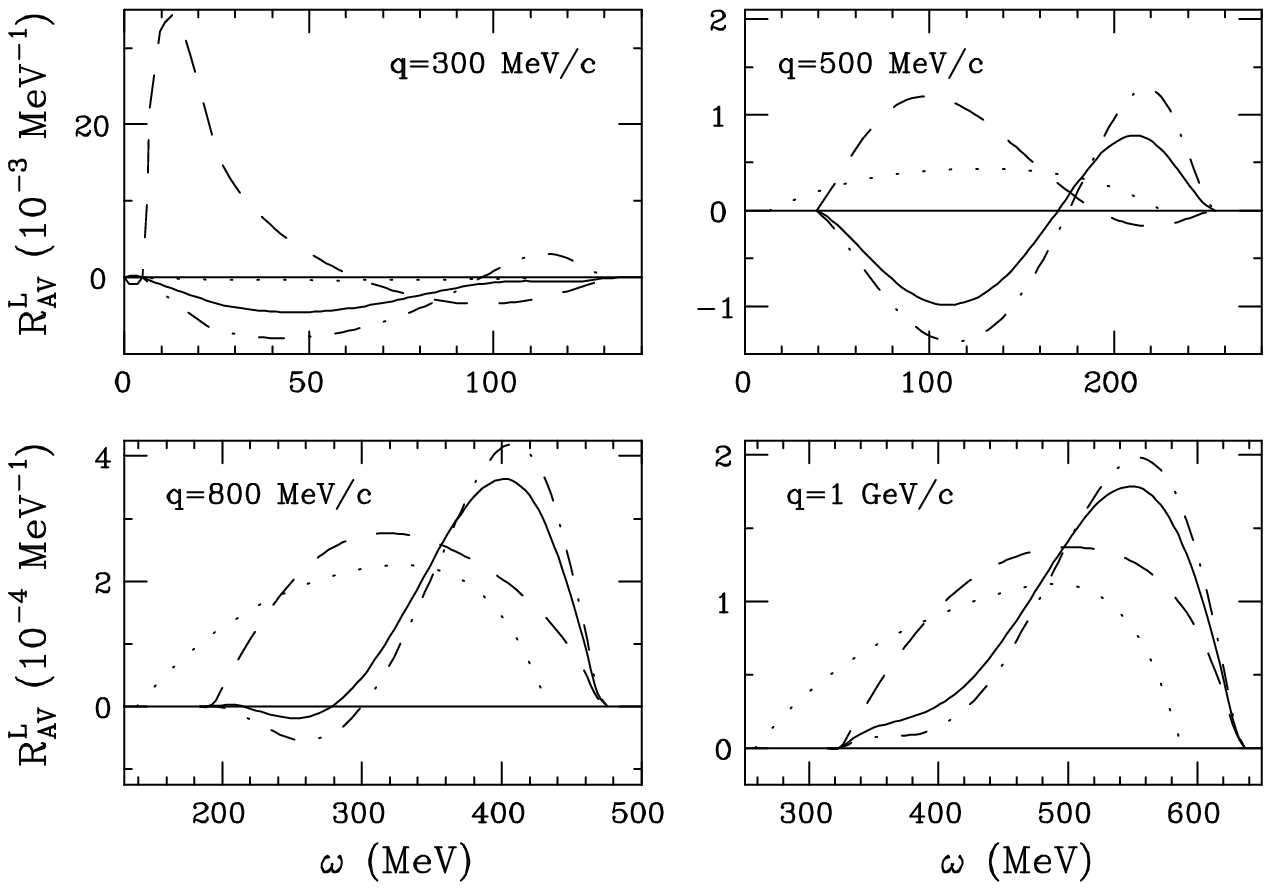,height=68mm}}
\vskip -3mm
\end{center}
\caption{
The pv longitudinal response versus $\omega$ at $q=$ 300, 500, 800 and
1000 MeV/c. The dotted lines correspond to the free RFG case ($k_F=$ 225
MeV/c),
the solid lines to the HF-RPA calculation ($k_F=$ 200 MeV/c);
the dashed lines represent the pure ring approximation, whereas the dot-dashed
ones the pure exchange contribution.
  }
\label{fig:RLAV-HF-ring-exch-rpa}
\end{figure}

The calculations displayed in the following~\cite{Bar95b} span a rather wide
range of transferred momenta. Hence, one has to cope with relativity and we
have
resorted to the relativistic Fermi gas (RFG) model, with inclusion of fully
antisymmetrized random phase approximation (RPA) correlations built on a HF
basis (from the discussion above, it should be clear that antisymmetrizing RPA
-- i.~e. including contributions such as those of Fig.~\ref{fig:diagrams-corr}c
-- is mandatory). As the input nucleon-nucleon interaction, we have chosen a
version of the Bonn potential~\cite{Mac87}, which accounts for $\pi$, $\rho$,
$\sigma$ and $\omega$ exchange. We refer the reader to Ref.~\cite{Bar95a} for a
thorough discussion of the model, together with its application to the
calculation of the em charge response.

A technical issue one should note here concerns the choice of the
Fermi momentum $k_F$, the only free parameter of the model. In the free Fermi
gas, $k_F$ is usually chosen to reproduce the experimental width of the
inclusive ($e,e'$) cross section; when correlations are introduced, the width
of
the response is modified and one should then change $k_F$ accordingly.
As shown in Ref.~\cite{Bar95a}, one should allow for a moderate reduction of
$k_F$: in $^{12}$C, for instance, one has $k_F=225$ MeV/c in the free model and
$k_F\approx200$ MeV/c in the correlated one. These are the values employed in
the following, together with $k_F=225$ MeV/c in the correlated model, which
should be adequate for medium-heavy nuclei and has the purpose of illustrating
the dependence on $k_F$.
\begin{figure}[t]
\begin{center}
\mbox{\epsfig{file=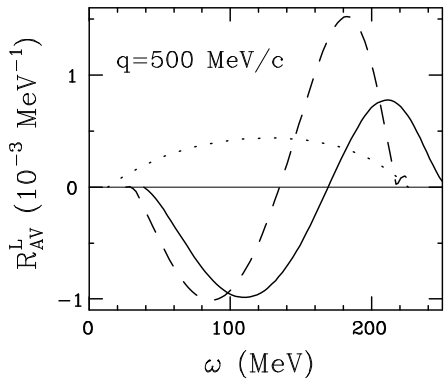,height=50mm}}
\vskip -3mm
\end{center}
\caption{
The pv longitudinal response is shown as a function of $\omega$ at
$q=$ 500 MeV/c. The dotted curve corresponds to the free RFG case, the solid
curve to the HF-RPA approximation with the full Bonn potential, the dashed
curve
is obtained in the HF-RPA approximation with a pure pion-exchange interaction;
$k_F=$ 225 MeV/c for the free RFG and $k_F=$ 200 MeV/c for the correlated
response.
  }
\label{fig:RLAV-pion}
\end{figure}

The results~\cite{Bar95b} for $R^L_{AV}$ are displayed in
Fig.~\ref{fig:RLAV-HF-RPA} at various momentum transfers. At moderate momenta
one can see the huge effect induced by correlations, which stays sizable even
at
1~GeV/c. Note the oscillating behaviour at $q=500$ MeV/c.
As discussed above, not all the correlations contribute equally to this
outcome.
HF correlations (Fig.~\ref{fig:diagrams-corr}a) are essentially filtered out,
apart from some hardening of the response already observed in
$R^L$~\cite{Bar95a}, and most of the effect enters through RPA.

It is interesting to see how different mesons contribute to RPA: in
Fig.~\ref{fig:RLAV-HF-ring-exch-rpa} the pure ring (direct) and the pure
exchange approximations for $R^L_{AV}$ are compared to the full HF-RPA
calculation. Both the ring diagrams --- due solely to the $\sigma$ and
$\omega$,
--- and the exchange ones --- dominated by the pion (the $\rho$ giving a small
contribution), --- have a strong impact on $R^L_{AV}$: however, the full
calculation turns out to be rather close to the approximation including only
exchange terms. What happens is that the {\em interference} of the pion with
the
$\sigma$ and $\omega$ is washing out their direct contribution; hence, the
final
result is close to what one should get in a pure pionic model.
This outcome is illustrated in Fig.~\ref{fig:RLAV-pion}, where $R^L_{AV}$ is
displayed at $q=500$ MeV/c in the full and pion-only models.

Note that in Ref.~\cite{Alb93} it had been shown that calculations in the
pionic
model can be adequately performed, over a wide range of momenta, in first order
perturbation theory. It is rather amusing that, after the smoke has cleared,
all
the complexities introduced in the nuclear dynamics conspire to yield a pv
longitudinal response described by just one exchange diagram.

\subsection{ Strangeness
}
\label{subsec:strangeness}

\begin{figure}[t]
\begin{center}
\mbox{\epsfig{file=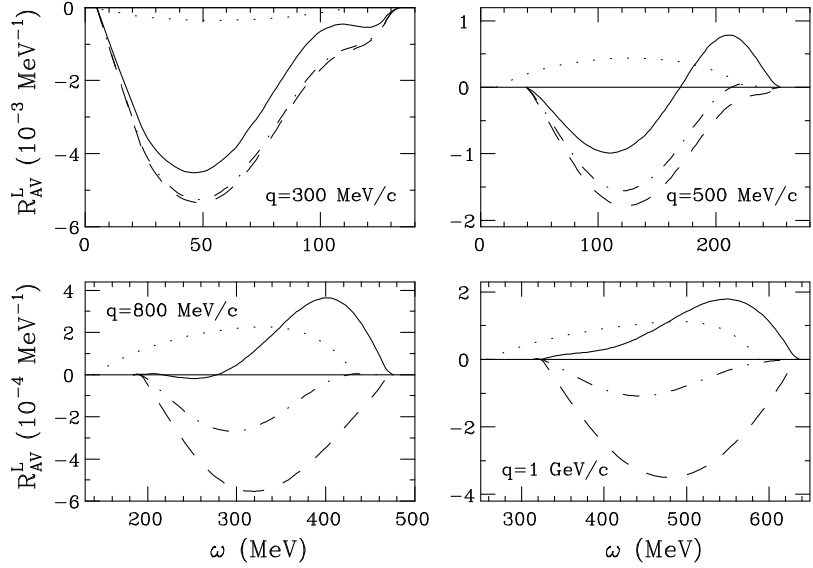,height=75mm}}
\vskip -3mm
\end{center}
\caption{
The pv longitudinal response is displayed versus $\omega$ at
$q=$ 300, 500, 800 and 1000 MeV/c.
The dotted curves refer to the free RFG case ($k_F=$ 225 MeV/c), the solid,
dot-dashed and dashed curves represent the HF-RPA results ($k_F=$ 200 MeV/c)
with $G_E^{(s)}$ given by the three models of Eq.~(\protect{\ref{eq:strange}}),
respectively.
  }
\label{fig:RLAV-strangeness}
\end{figure}

$R^L_{AV}$ is, roughly speaking, proportional to the weak-neutral electric form
factor $\widetilde G_E$. The latter, in turn, can be expressed as a combination
of the standard proton and neutron electric form factors, with weights dictated
by the standard model, plus a possible electric strange form factor,
$G^{(s)}_E$, induced by a nonzero strangeness content of the
nucleon.~\cite{Don92}

In order to test the sensitivity of $R^L_{AV}$ to variations in $G^{(s)}_E$,
one
has to choose a parametrization for the latter.
We have taken a Galster-like form:
\begin{equation}
  G_E^{(s)}(\tau) = \rho_s {\tau\over \left[1+\lambda_D^V\tau\right]^2}
    {1\over\left[1+\lambda_E^{(s)}\tau\right]} ,
\label{eq:Galster}
\end{equation}
with $\lambda_D^V=4.97$. Three ``reasonable'' choices for the parameters
$\rho_s$ and $\lambda_E^{(s)}$ have been considered in past
work~\cite{Don92,Bar94}, namely
\begin{equation}
\begin{array}{rcl}
  \rho_s &=& 0 \\
  \rho_s &=& -3\, ,\quad\lambda_E^{(s)}=5.6 \\
  \rho_s &=& -3\, , \quad\lambda_E^{(s)}=0 \,.
\label{eq:strange}
\end{array}
\end{equation}
They correspond, respectively, to the absence of strangeness, to the same
momentum dependence as the electric neutron form factor and to a pure
dipole form factor.

In Fig.~\ref{fig:RLAV-strangeness}, the results for $R^L_{AV}$ with these three
choices for $G^{(s)}_E$ are presented.~\cite{Bar95b} It is apparent that at
large momenta, where $G^{(s)}_E$ is not suppressed by the factor of $\tau$ in
the numerator of (\ref{eq:Galster}) (which is dictated by the fact that the
total strangeness of the nucleon is zero), $R^L_{AV}$ is significantly
sensitive
to the strength of $G^{(s)}_E$.

\section{ Disentangling nuclear and nucleonic physics}
\label{sec:disent}

\begin{figure}[p]
\mbox{\epsfig{file=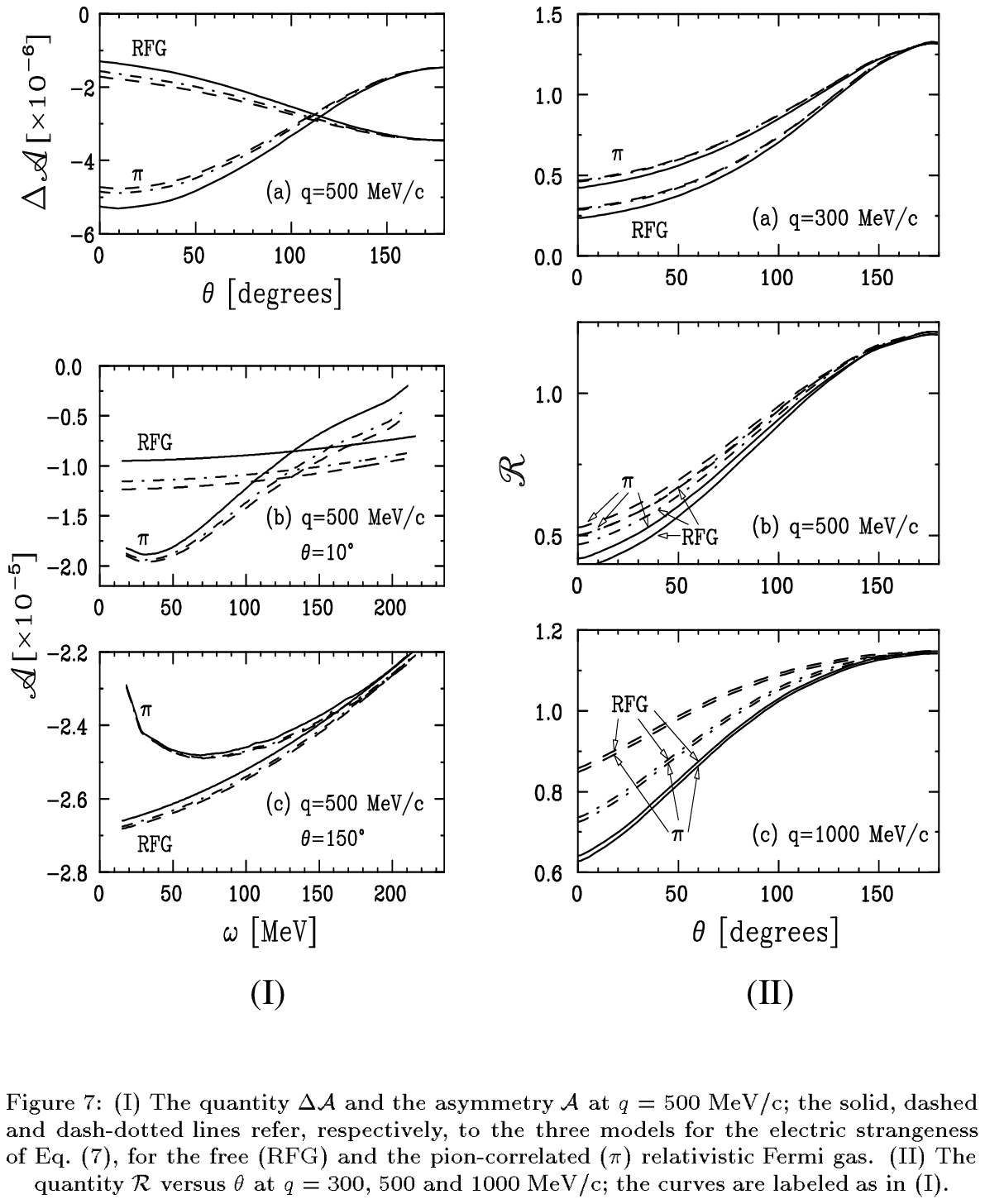}}
\label{fig:delta-r_es_500}
\end{figure}

In view of the results discussed in Sec.~\ref{sec:w-nlr}, one may obviously
wonder whether it is possible to disentangle the sensitivities to the nuclear
dynamics and to the nucleonic structure.
This is, of course, a general problem and concerns all the response functions
measurable in polarized electron scattering. Furthermore, since, given the
present experimental capabilities, there is no way to separate all the
responses, one has to tackle this problem through the asymmetry
(\ref{eq:asymmetry}). A general procedure, within the RFG scheme, has been
devised in Ref.~\cite{Bar94} and there applied to the pure pionic model.
Calculations in the full meson-exchange model are still in progress; however,
pionic dynamics has been shown in Sec.~\ref{sec:w-nlr} to be adequate for the
pv
longitudinal channel, whereas correlations in the other channels do not have
the
same disruptive effects and, moreover, act in the same way in the response
functions entering the numerator and the denominator of
Eq.~(\ref{eq:asymmetry}), thus making the asymmetry an observable much less
sensitive to transverse and axial correlations than the responses themselves.
For these reasons, the results based on the pionic model should be adequate for
the purpose of illustrating the procedure.

To enhance the effect of nuclear correlations, one can introduce the following
integrated (over the transferred energy) observable:
\begin{equation}
  \Delta{\cal{A}}(q,\theta) \equiv {1 \over{\Delta\omega}}\Bigl[
  \int_{\omega_{min}}^{\omega_{QEP}}\!\!d\omega
   {\cal{A}}(\theta;q,\omega) -
  \int_{\omega_{QEP}}^{\omega_{max}}\!\!d\omega
   {\cal{A}}(\theta;q,\omega)\Bigr]\ ,
\label{eq:delta_a}
\end{equation}
where $\omega_{min}$ and $\omega_{max}$ represent the boundaries of the
response
region and $\omega_{QEP}$ the quasielastic peak (QEP) energy.
Since a sizable fraction of the contribution induced by correlations is
antisymmetric around the QEP, whereas variations of the nucleon form factors
generate a uniform shift of the asymmetry, $\Delta{\cal{A}}$ turns out to be
rather unsensitive to modifications of the nucleon structure, as one can see in
Fig.~\ref{fig:delta-r_es_500}I, where, for the sake of illustration, the
sensitivity of $\Delta{\cal{A}}$ to the electric strangeness content of the
nucleon is displayed.

Constructing an observable desensitized to the nuclear dynamics is slightly
more
involved and requires sum rule considerations. In fact, in a relativistic
description of nuclear dynamics the nucleon form factors cannot be simply
factored out: however, in the RFG one can define {\em reduced} response
functions --- for which the factorization is approximately true --- satisfying
the sum rule:
\begin{equation}
  S^\alpha(q,\omega) = v_\alpha R^\alpha(q,\omega)/X'_\alpha,
\end{equation}
with the dividing factors $X'_\alpha$ given in Ref.~\cite{Bar94}.
One can then define the following observable:
\begin{equation}
  {\cal R}(q,\theta) \equiv \frac{
    \int_{\omega_{min}}^{\omega_{max}}\!\!d\omega\ W^{PV}(q,\omega)
    \big/\widetilde{X}'_T }{
    \int_{\omega_{min}}^{\omega_{max}}\!\!d\omega\ W^{EM}(q,\omega)
    \big/X'_T } \, ,
\label{eq:R}
\end{equation}
with $W^{PV} = v_L R^L_{AV} +v_T R^T_{AV} +v_{T'}R^{T'}_{VA}$ and
$W^{EM} = v_L R^L +v_T R^T$. ${\cal R}$ is displayed in
Fig.~\ref{fig:delta-r_es_500}II, with $G^{(s)}_E$ given by the three models of
Eq.~(\ref{eq:strange}). At high momenta, where the strange form factor is not
suppressed by its momentum dependence, ${\cal R}$ shows a remarkable
sensitivity to $G^{(s)}_E$, while being practically independent of nuclear
correlations.

Of course, a large amount of work has to be done to sophisticate the nuclear
model (accounting, e.~g., for meson-exchange currents, short range
correlations,
2p-2h, ...) and to test the model dependence of the observables discussed above
(on the input nucleon-nucleon interaction, on finite size effects, ...).
These preliminary results look, however, promising enough to let one hope that
quasielastic scattering may be included in the landscape of future pv
experiments.

\vfill
\eject

\end{document}